\documentstyle[12pt,epsfig]{article}
\voffset     = - 0.10 in
\begin{document}
\thispagestyle{empty}

%\hfill {\Large\bf DRAFT}

%\hfill {\Large\bf \today }

\vspace*{1.0 in}

\begin{center}
{\Large\bf
Non-compensation of an Electromagnetic Compartment
of a Combined Calorimeter 
}
\end{center}

\bigskip
\bigskip

\begin{center}
{\large\bf Y.A.\ Kulchitsky,  M.V.\ Kuzmin}

\smallskip

{\sl Institute of Physics,  National Academy of Sciences,  Minsk,  Belarus}\\

\smallskip

\& \ \ 
{\sl JINR,  Dubna,  Russia}

\bigskip

{\large\bf V.B.\ Vinogradov}

\smallskip
{\sl JINR,  Dubna,  Russia}
\bigskip

\end{center}

\vspace*{\fill}

\begin{abstract}
The method of extraction of the $e/h$ ratio, the degree of non-compensation, 
of the electromagnetic compartment of the combined 
calorimeter is suggested.
The $e/h$ ratio of $1.74\pm0.04$ has been determined
on the basis of the 1996 combined calorimeter test beam data. 
This value agrees with the prediction that $e/h > 1.7$ for this 
electromagnetic calorimeter.
\end{abstract}

\newpage

%%%%%%%%%%%%%%%%%%%%%
\section{Introduction}
%%%%%%%%%%%%%%%%%%%%%
\hspace{6mm}
The existing calorimetric complexes (CDF,  D0,  H1 etc.) as well as the
the future huge ones (ATLAS
%1 
\cite{atcol94},  CMS etc.) 
at the CERN Large Hadron Collider (LHC)
are the combined calorimeters
with the electromagnetic and hadronic compartments.
For the energy reconstruction and description of the longitudinal
development of  a hadronic shower it is necessary to know the $e / h$ ratios,  
the degree of non-compensation,    of these  calorimeters. 
As to the ATLAS Tile barrel calorimeter there is
the detailed information about the $e / h$ ratio presented in
%2
\cite{ariztizabal94},
%3 
\cite{juste95},
%4 
\cite{budagov96-72},
%5 
\cite{TILECAL96},
%6  
\cite{kulchitsky99}.
But as to the liquid argon electromagnetic calorimeter such information
practically absent.

The aim of the present work is to develop the  method and to determine
the value of the $e / h$ ratio of the LAr electromagnetic compartment. 

This work has been performed on the basis of 
the 1996 combined test beam data 
%7
\cite{cobal98}.
Data were taken on the H8 beam of the CERN SPS, with pion and
electron beams of  10, 20, 40, 50, 80, 100, 150 and 300 GeV/c.

%%%%%%%%%%%%%%%%%%%%%%%%%%%%%%%%%%%%%%%%%%%%
\section{The Combined Prototype Calorimeter}
%%%%%%%%%%%%%%%%%%%%%%%%%%%%%%%%%%%%%%%%%%%%
\hspace{6mm}
The future ATLAS experiment 
\cite{atcol94} 
will include in the central (``barrel'') region a calorimeter system composed
of two separate units: 
the liquid argon electromagnetic calorimeter (LAr)
%8
\cite{LARG96}
and the tile iron-scintillating hadronic calorimeter (Tile) 
\cite{TILECAL96}. 

\begin{figure}[tbph]
\begin{center}
\mbox{\epsfig{figure=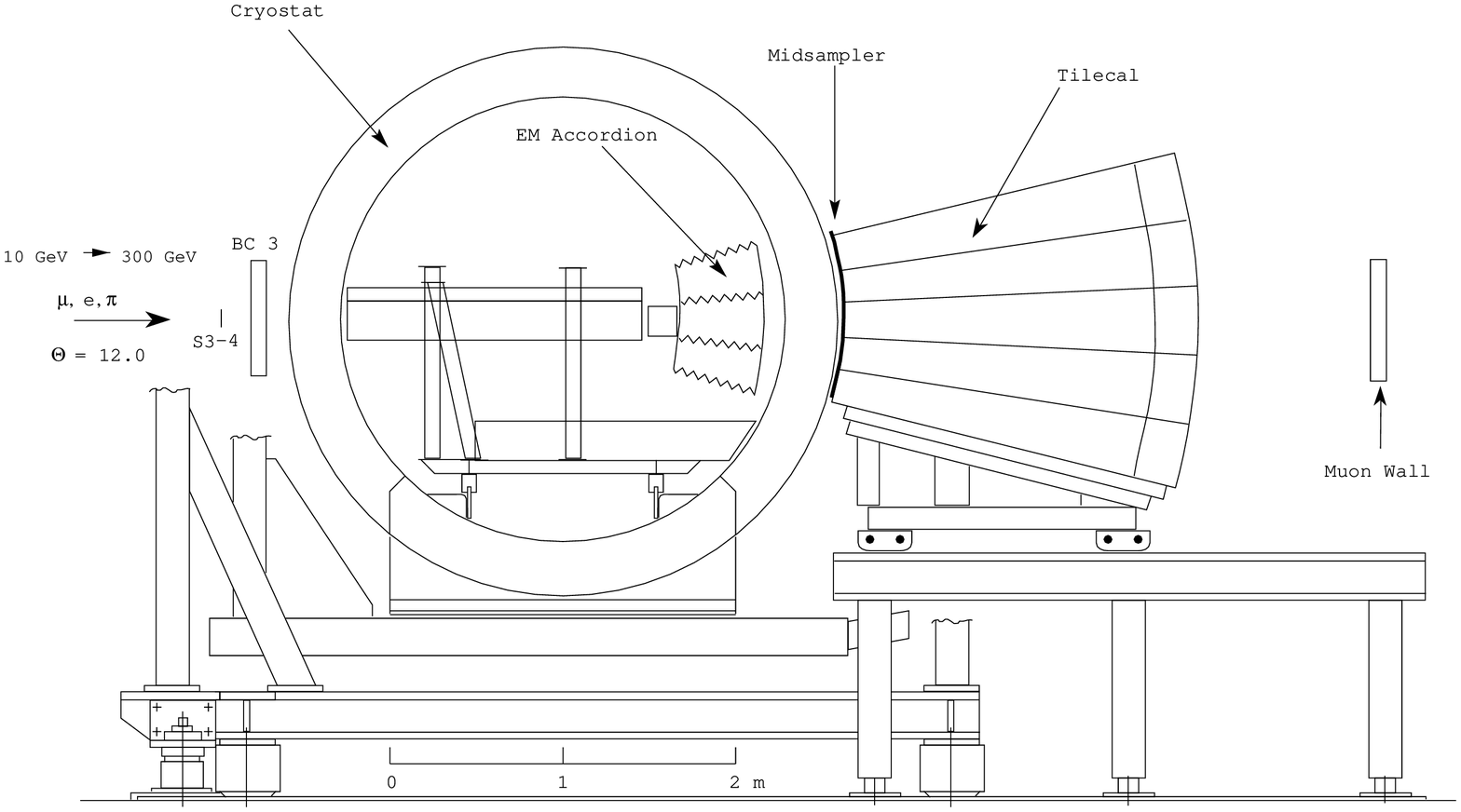,width=0.95\textwidth,height=0.40\textheight}} 
\end{center}
 \caption{ 
   Test beam setup for the ATLAS combined prototype calorimeter.}
\label{fv1}
\end{figure}

For detailed understanding of performance of the future ATLAS
combined calorimeter the combined calorimeter prototype setup
has been made consisting of the LAr electromagnetic calorimeter prototype
inside the cryostat and downstream the Tile calorimeter prototype
as shown in Fig.~\ref{fv1}. 

The dead material between the two calorimeters was about
$2.2\ X_0$ or $0.28\ \lambda_I^\pi$.
Early showers in the liquid argon were kept to a minimum by placing
the light foam material in the cryostat upstream of the calorimeter. 

The two calorimeters have been placed with their central axes at an angle 
to the beam of $12^\circ$.
At this angle the two calorimeters have an active thickness of 
10.3 $\lambda_I$.

Between the active part of the LAr and the Tile detectors
a layer of scintillator was installed,  called the midsampler. 
The midsampler consists of five scintillators,  $20\times100$\ cm$^2$ each, 
fastened directly to the front face of the Tile modules.
The scintillator is 1 cm thick.

Beam quality and geometry were monitored with a set 
of beam wire chambers BC1,  BC2,  BC3
and trigger hodoscopes placed upstream of the LAr cryostat.

To detect punchthrough particles and to
measure the effect of longitudinal leakage
a ``muon wall'' consisting of 10 scintillator counters (each 2 cm thick)
was located behind the calorimeters at a distance of about 1 metre.

%%%%%%%%%%%%%%%%%%%%%%%%%%%%%%%%%%%%%%%%%%%%%%%%%%%%%%%%%
\subsection{The Electromagnetic Liquid Argon Calorimeter}
%%%%%%%%%%%%%%%%%%%%%%%%%%%%%%%%%%%%%%%%%%%%%%%%%%%%%%%%%
\hspace{6mm}
The electromagnetic LAr calorimeter  prototype
consists of a stack of three azimuthal modules,  
each one spanning $9^\circ$ in
azimuth and extending  over 2 m along the Z direction.
The calorimeter structure is defined by
2.2 mm thick steel-plated lead absorbers,  folded to an accordion shape and
separated by 3.8 mm gaps,  filled with liquid argon.
The signals are collected by Kapton electrodes located in the gaps.
The calorimeter extends from
an inner radius of 131.5 cm to an outer radius of 182.6 cm, 
representing (at $\eta = 0$) a total
of 25 radiation lengths ($X_0$),  or 1.22 interaction lengths
($\lambda_I$) for protons.
The calorimeter is longitudinally
segmented into three compartments of
$9\ X_0$,  $9\ X_0$ and $7\ X_0$,  respectively.
More details about this prototype can be found in 
\cite{atcol94},
%9
\cite{ccARGON}.

In front of the EM calorimeter a presampler was mounted. 
The active depth of liquid argon in the presampler is 10 mm and
the strip spacing is 3.9 mm.

The cryostat has a cylindrical form
with 2 m internal diameter,  filled with liquid argon,  and
is made out of a  8 mm thick inner stainless-steel vessel, 
isolated by 30 cm of low-density foam (Rohacell),  itself protected by a
1.2 mm thick aluminum outer wall.

%%%%%%%%%%%%%%%%%%%%%%%%%%%%%%%%%%%%%%%%%%
\subsection{The Hadronic Tile Calorimeter}
%%%%%%%%%%%%%%%%%%%%%%%%%%%%%%%%%%%%%%%%%%
\hspace{6mm}
The  hadronic Tile calorimeter 
is a sampling device using
steel as the absorber and scintillating tiles as the active material 
\cite{TILECAL96}.
The innovative feature of the design is the orientation of the tiles
which are placed in planes perpendicular to the Z direction 
%10
\cite{gild91}.
For a better
sampling homogeneity the 3 mm thick scintillators are staggered in the
radial direction.
The tiles are separated along  Z  by 14 mm of steel,  giving a
steel/scintillator volume ratio of 4.7.
Wavelength shifting fibers (WLS) running radially collect light from the
tiles at both of their open edges.
The hadron calorimeter prototype consists of an azimuthal
stack of five modules.
Each module covers $2\pi/64$ in azimuth and extends
1 m along the Z direction,  such that the front face covers
$100\times20$\ cm$^2$. 
The radial depth, 
from an inner radius of 200 cm to an outer radius of 380 cm, 
accounts for 8.9 $\lambda$ at $\eta = 0$\  (80.5 $X_0$).
Read-out cells are defined by grouping together a bundle of fibers
into one photomultiplier (PMT).
Each of the 100 cells is read out by two PMTs and is fully
projective in azimuth (with $\Delta \phi = 2\pi/64 \approx 0.1$), 
while the segmentation along the Z
axis is made by grouping fibers into read-out cells spanning
$\Delta Z = 20$ cm ($\Delta \eta \approx 0.1$) and is therefore not projective
Each module is read out in four longitudinal segments
(corresponding to about 1.5,  2,  2.5 and 3 $\lambda_I$ at $\eta = 0$).
More details of this prototype can be found in
\cite{atcol94},
\cite{ariztizabal94}.

%%%%%%%%%%%%%%%%%%%%%%%%
\section{Event Selection}
%%%%%%%%%%%%%%%%%%%%%%%%
\hspace{6mm} 
We applied some similar to 
\cite{cobal98} 
cuts to eliminate the non-single track pion events, the beam halo,  
the events with an interaction before LAr calorimeter,  the events with the 
longitudinal leakage, the electron  and muon events.
The set of cuts  is the following:
\begin{itemize}
\item
the single-track pion events were selected by requiring the pulse height 
of the beam scintillation counters  and the energy released in the 
presampler of the electromagnetic calorimeter to be compatible with that 
for a single particle;
\item
the beam halo events were removed with appropriate cuts on the horizontal
and vertical positions of the incoming track impact point and the space 
angle with respect to the beam axis as measured with the beam chambers;
\item
the electron events were removed by the requirement that the energy deposited 
in the LAr calorimeter is less than 90 \% of the beam energy;
\item 
a cut on the total energy rejects incoming muon;
\item
the events with the obvious longitudinal leakage were removed by requiring of 
no signal from the punchthrough particles in the muon walls;
\item
to select the events with the hadronic shower origins in the first sampling 
of the LAr calorimeter; events with the energy depositions in this sampling
compatible with that of a single minimum ionization particle were rejected;
\item  
to select the events with the well developed hadronic showers energy depositions 
were required to be more than 10 \% of the beam energy in the 
electromagnetic calorimeter and less than 70 \% in the hadronic calorimeter. 
\end{itemize}

%%%%%%%%%%%%%%%%%%%%%%%%%%%%%%%%%%%%%%%%%%%
\section{The $e/h$ ratio of  the LAr Electromagnetic Compartment}
%%%%%%%%%%%%%%%%%%%%%%%%%%%%%%%%%%%%%%%%%%%
\hspace{6mm}
The response,  $R_h$,  of a calorimeter to a hadronic shower is the sum of 
the contributions  from the electromagnetic,  $E_e$,  and hadronic,  $E_h$,  
parts of the incident energy
%11
\cite{groom89}
\begin{equation}
	E = E_e + E_h \ , 
\label{ev18}
\end{equation}
\begin{equation}
	R_h = 
	e \cdot E_e + h \cdot E_h = 
	e \cdot E \cdot (f_{\pi^o} +(h / e) \cdot (1 - f_{\pi^o})) \ ,  
\label{ev9}
\end{equation}
where $e$ ($h$) is the energy independent coefficient of
transformation of the electromagnetic 
(pure hadronic,  low-energy hadronic activity) energy to response, 
$f_{\pi^o} = E_e / E$ is the fraction of electromagnetic energy.
From this
\begin{equation}
	E = \frac{e}{\pi} \cdot \frac{1}{e} \cdot R_h \ , 
\label{ev16}
\end{equation}
where
\begin{equation}
	\frac{e}{\pi} = \frac{e/h}{1 + (e/h-1) f_{\pi^o}} \ . 
\label{ev10}
\end{equation}

In the case of the combined calorimeter the incident beam energy,  
						$E_{beam}$,  
is deposited into  the LAr compartment,  	$E_{LAr}$, 
into Tilecal compartment,  			$E_{Tile}$, 
and into the dead material
between the LAr and Tile calorimeters,  	$E_{dm}$, 
\begin{equation}
	E_{beam} = E_{LAr} + E_{Tile} + E_{dm} \ .
\label{ev13}
\end{equation}

Using relation (\ref{ev16}) the following expression has been obtained:
\begin{equation}
	E_{beam} = 
		  c_{LAr}  \cdot 
		\Bigl( \frac{e}{\pi} \Bigr)_{LAr}  \cdot R_{LAr} 
		+ c_{Tile} \cdot 
		\Bigl( \frac{e}{\pi} \Bigr)_{Tile} \cdot R_{Tile} 
        	+ E_{dm} \ , 
\label{ev7}
\end{equation}
where $c_{LAr} = 1 / e_{LAr}$ and $c_{Tile} = 1 / e_{Tile}$. 
From this expression the value of the $(e / \pi)_{LAr}$ ratio can be obtained
\begin{equation}
	\Bigl( \frac{e}{\pi} \Bigr)_{LAr} =
	\frac{E_{beam} - E_{Tile}  - E_{dm}}{c_{LAr} \cdot R_{LAr}} \ , 
\label{ev1}
\end{equation}
where
\begin{equation}
	E_{Tile} = 
	c_{Tile} \cdot \Bigl( \frac{e}{\pi} \Bigr)_{Tile} \cdot R_{Tile}
\label{ev15}
\end{equation}
is the energy released in the Tile calorimeter.

The $(e/h)_{LAr}$ ratio  and 
\begin{equation}
	f_{\pi^o, LAr} = k_{LAr} \cdot ln{E_{beam}} 
\label{efpl}
\end{equation} 
can be inferred from the energy dependent $(e/\pi)_{LAr}$ ratios:
\begin{equation}
	\Bigl( \frac{e}{\pi} \Bigr)_{LAr} = 
		\frac{(e/h)_{LAr}}{1 + ((e/h)_{LAr} - 1) f_{\pi^o, LAr}} \ .
\label{ev5}
\end{equation} 

We used the value $(e/h)_{Tile}=1.3$ 
\cite{budagov96-72}
and the following  expression for the electromagnetic fraction of a hadronic 
shower in the Tilecal calorimeter
\begin{equation}
	f_{\pi^o, Tile} = k_{Tile} \cdot ln E_{Tile}.
\label{ev3}
\end{equation}
with
$k_{Tile} = 0.11$
%12  
\cite{wigmans91},
%13  
\cite{acosta92}.
 
For the $c_{LAr}$ constant the value of 1.1,  obtained in
%14 
\cite{kulchitsky98-336},  
\cite{cobal98},  
was used.

The algorithm for finding the $c_{Tile}$ and $c_{dm}$ constants will be considered
in the next section.

%%%%%%%%%%%%%%%%%%%%%%%%%%%%%%%%
\section{The $c_{Tile}$ Constant}
%%%%%%%%%%%%%%%%%%%%%%%%%%%%%%%% 
\hspace{6mm}
For the determining of the $c_{Tile}$ constant the following procedure was 
applied.   
We selected the events which start to shower only in the hadronic
calorimeter.
To select these events the energies deposited 
in each sampling of the LAr calorimeter and in the midsampler 
are required to be compatible with that of a beam particle.   
We used the following expression for the normalized hadronic response 
%11
\cite{groom89}
\begin{equation}
	\frac{R_{Tile}^{c}}{E_{beam}} =
		\frac{c_{Tile}}{(e/h)_{Tile}}
		 ( 1 + \Bigl( \Bigl( \frac{e}{h} \Bigr)_{Tile} - 1 \Bigr) 
		\cdot {(f\pi^0)}_{Tile} ) \ , 
\label{ev14}
\end{equation}
where
\begin{equation}
	R_{Tile}^{c} = 
		R_{Tile} + \frac{c_{LAr}}{c_{Tile}} \cdot R_{LAr}
\label{ev17}
\end{equation}
is the Tile calorimeter response corrected on the energy loss in the
LAr calori\-meter, $f_{\pi^0, Tile}$ is determined by the formula (\ref{ev3}).

\begin{figure*}[tbph]
\begin{center}   
\begin{tabular}{c}
\mbox{\epsfig{figure=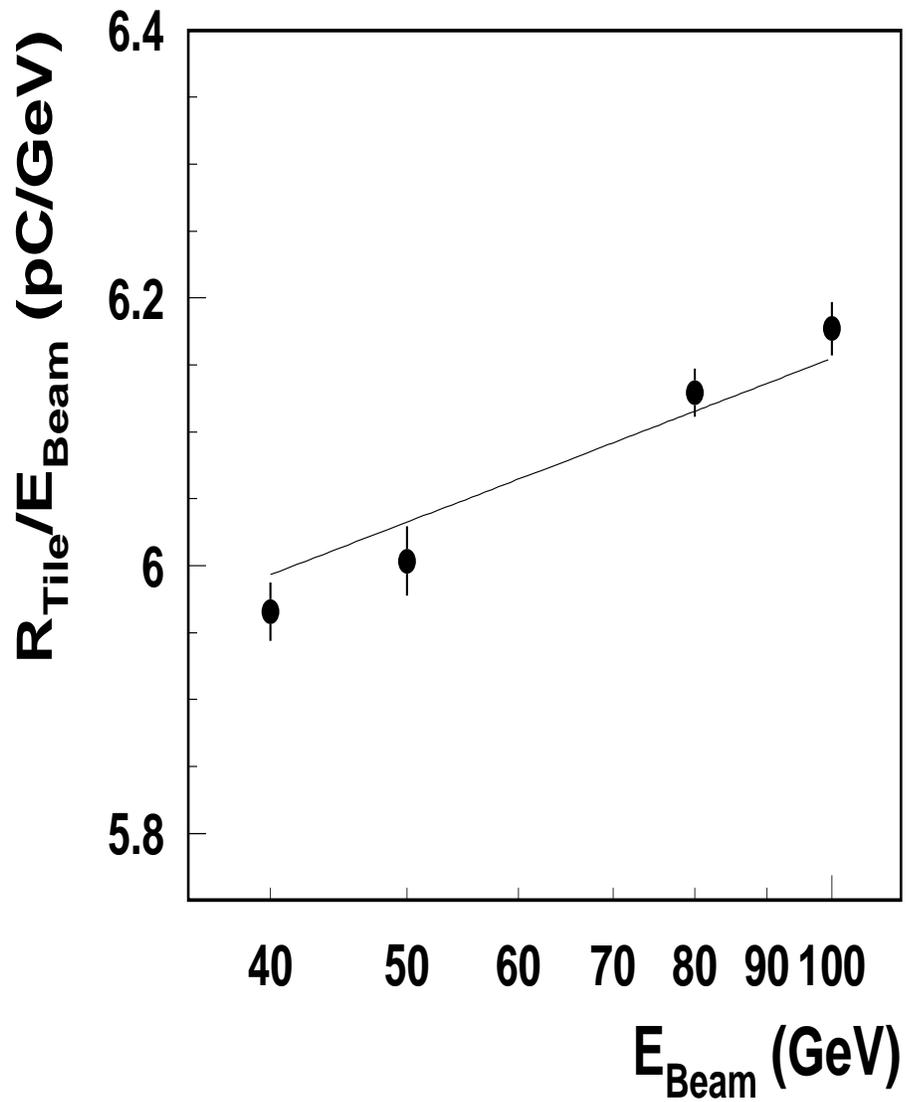,width=0.95\textwidth,height=0.9\textheight}} 
\\
\end{tabular}
\end{center}
       \caption{ 
	The corrected $R_{Tile}$ response as a function of the beam energy.
       \label{fv15}}
\end{figure*}

The values of $R_{Tile}^{c}$ are shown in Fig.\ \ref{fv15} together 
with the fitting line.
The obtained value of $c_{Tile}$ is equal to $0.145\pm0.002$.

%%%%%%%%%%%%%%%%%%%%%%%%%%%%%%%%%%%%%%%%%%%
\section{The Energy Loss in the Dead Material}
\begin{figure*}[tbph]
\begin{center}   
\begin{tabular}{c}
\epsfig{figure=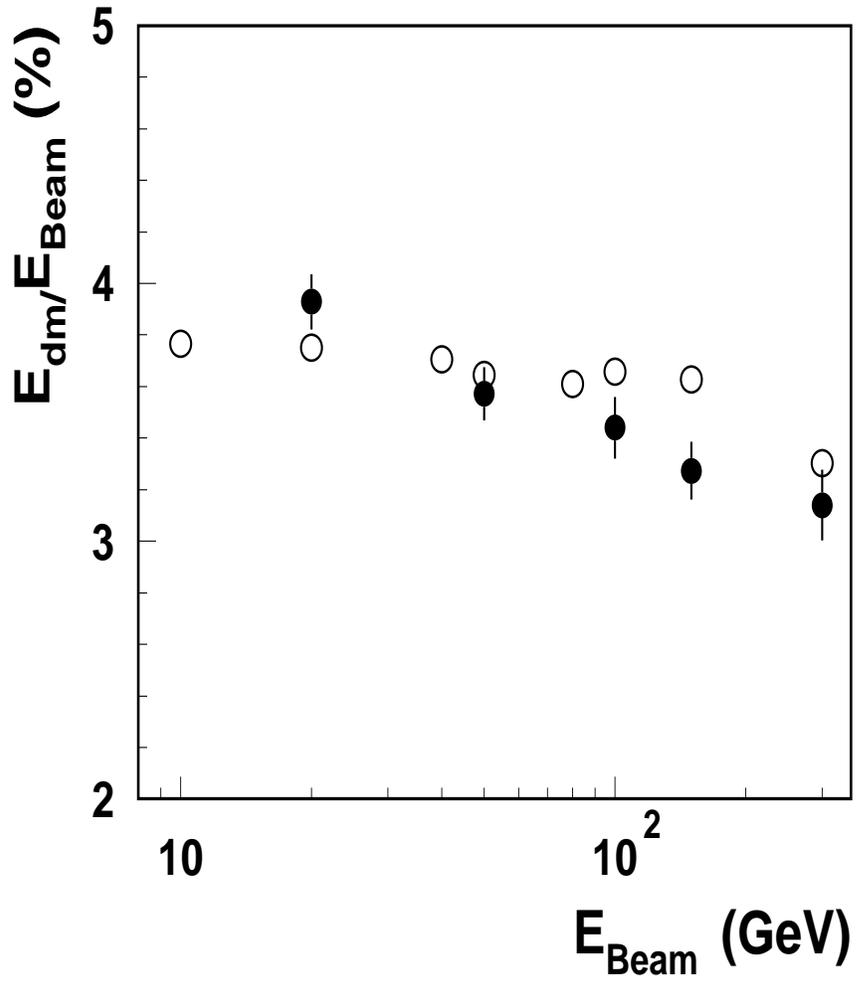,width=0.95\textwidth,height=0.80\textheight}
\\
\end{tabular}
\end{center}
       \caption{
	The comparison between the Monte Carlo simulation (solid circles)
	and the calculated values (open circles) for the average relative 
	energy losses in the dead material, $E_{dm}/E_{beam}$,  as a 
	function of the beam energy. 
       \label{f002}}
\end{figure*}

\begin{figure*}[tbph]
\begin{center}   
\begin{tabular}{c}
\epsfig{figure=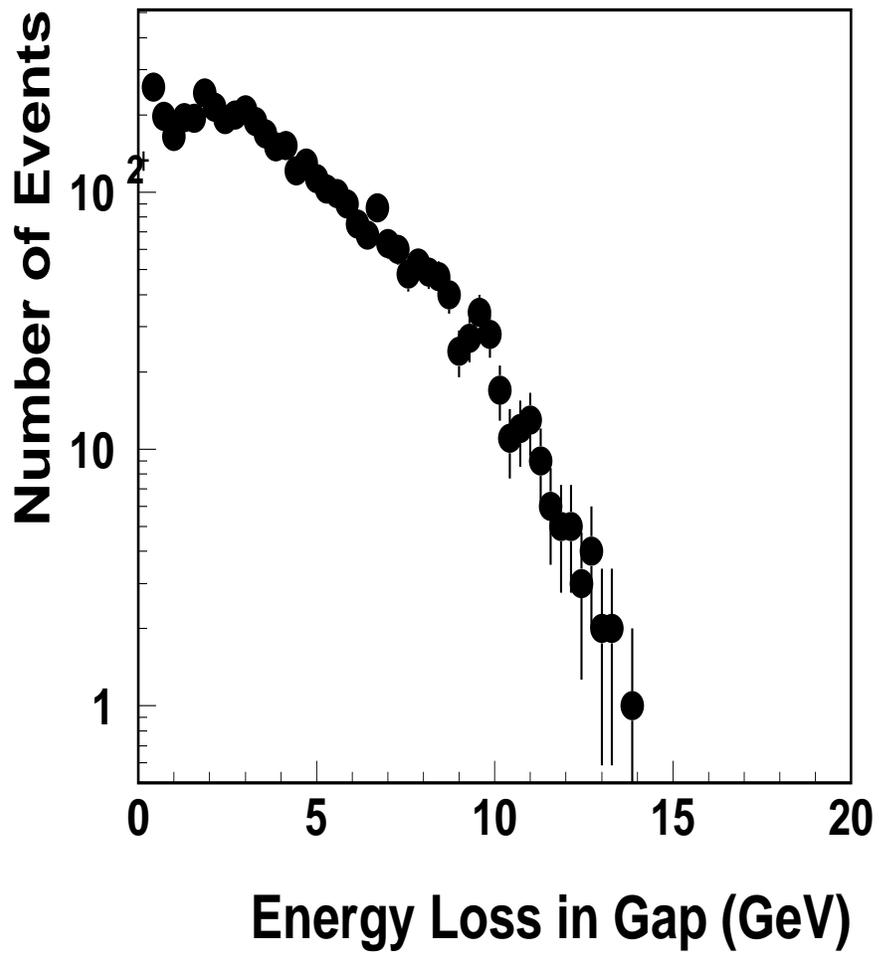,width=0.95\textwidth,height=0.80\textheight}
\\
\end{tabular}
\end{center}
       \caption{The distribution of energy loss in the dead material for 50 
$GeV$ pion beam.
       \label{f0021}}
\end{figure*}

%%%%%%%%%%%%%%%%%%%%%%%%%%%%%%%%%%%%%%%%%%%
\hspace{6mm}
Special attention has been devoted to understanding of
the energy loss in the dead material
placed between the active part of the LAr and the Tile detectors.
The term,  which accounts for
the energy loss in the dead material  between the LAr and Tile
calorimeters,  $E_{dm}$,   
is taken to be proportional to the geometrical mean 
of the energy released in the last electromagnetic compartment
($E_{LAr, 3}$) and the first hadronic compartment ($E_{Tile, 1}$)
\begin{equation}
\label{ev19}
	E_{dm} = 
		c_{dm} \cdot \sqrt{E_{LAr, 3} \cdot E_{Tile, 1}} 
\end{equation}
similar to
\cite{cobal98},
%15
\cite{combined94}.
The validity of this approximation has been tested by the Monte Carlo 
simulation and by the study of the correlation between the energy released
in the midsampler and the cryostat energy deposition 
\cite{cobal98},
%16 
\cite{bosman99},
%17  
\cite{atcol99}. 
We used the value of $c_{dm} = 0.31$.
This value has been obtained on the basis of the results of 
the Monte Carlo simulation performed by I.~Efthymiopoulos
%18
\cite{Efthymiopoulos}.
These Monte Carlo (Fluka) results (solid circles)
are shown in  Fig.~\ref{f002} together with the values (open  circles) 
obtained by using the expression (\ref{ev19}).
The reasonable agreement is observed.
The average energy loss in the dead material is equal to about $3.7 \%$. 
The typical distribution of the energy losses in the dead material between 
the LAr and Tile calorimeters for the real events at the beam energy of 
50 $GeV$,   obtained  by using Eq.\ \ref{ev19}, 
is shown  in Fig.~\ref{f0021}.

%%%%%%%%%%%%%%%%%%%%%%%%%%%%%%%%%%%%%%%%%%%%%%%%%%%%%%%
\section{The $(e/\pi)_{LAr}$ and $(e/h)_{LAr}$ Ratios.}
\begin{figure*}[tbph]
\begin{center}   
\begin{tabular}{cc}
\mbox{\epsfig{figure=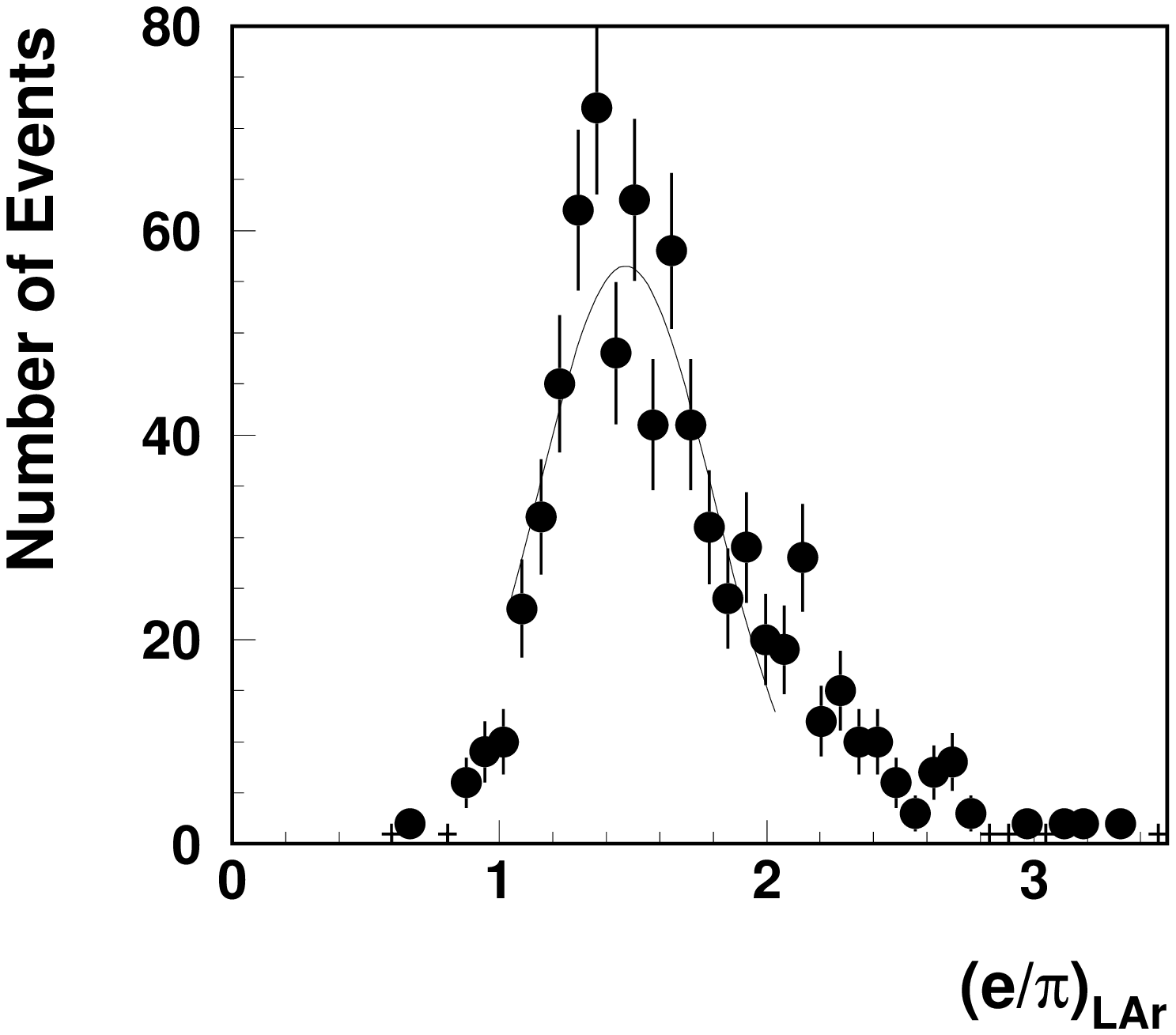,width=0.45\textwidth,height=0.4\textheight}} & 
\mbox{\epsfig{figure=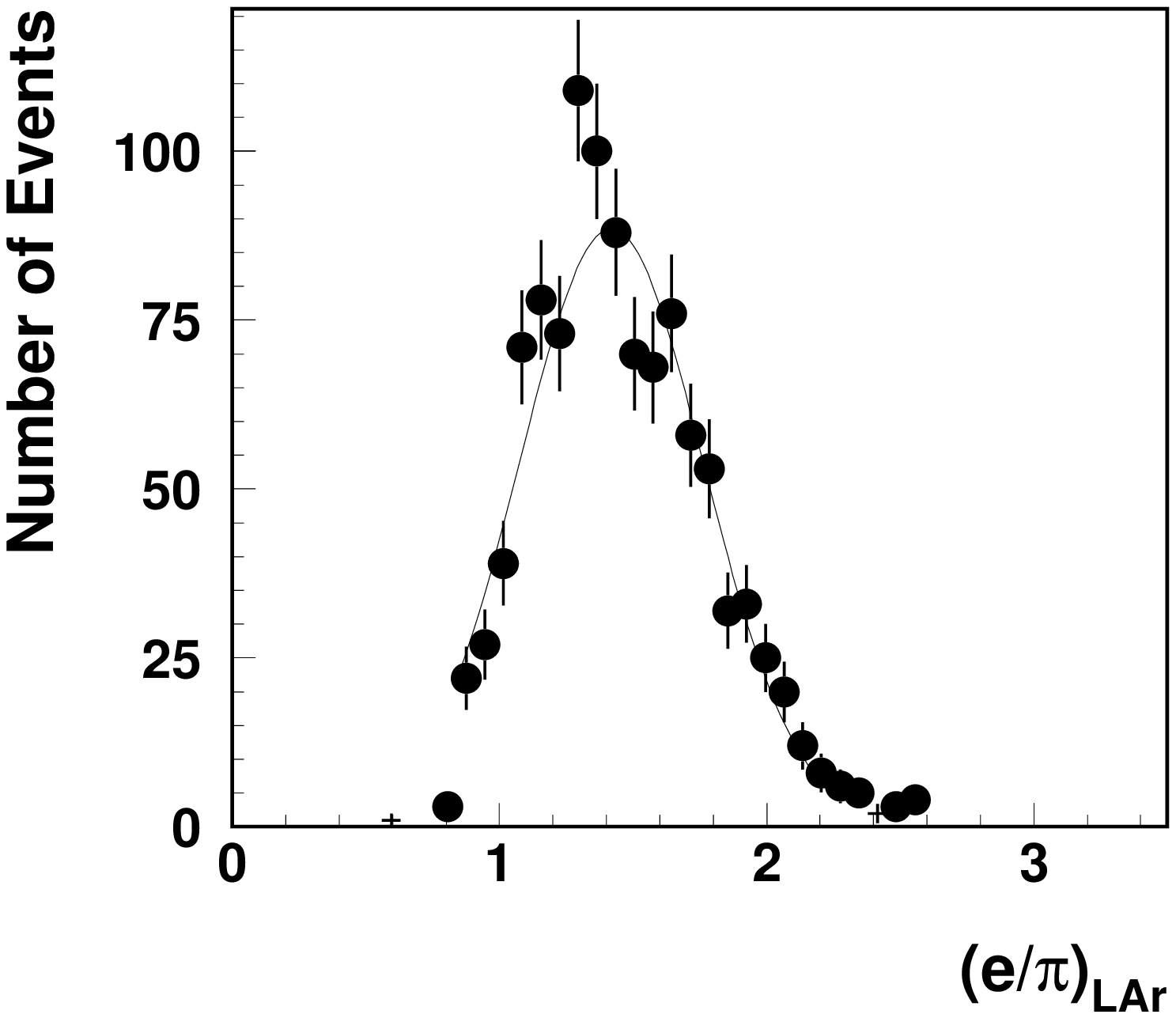,width=0.45\textwidth,height=0.4\textheight}} 
\\
\mbox{\epsfig{figure=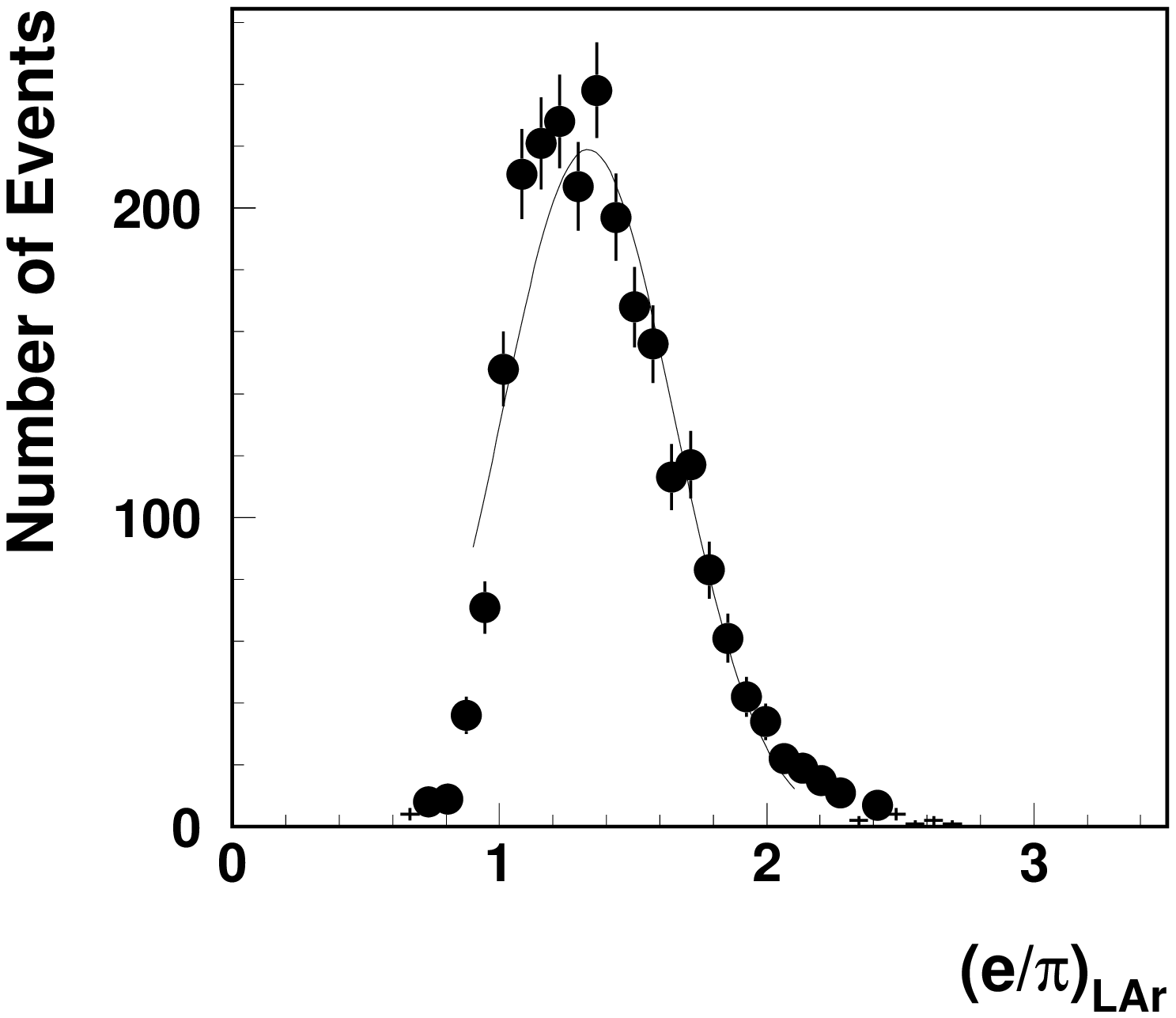,width=0.45\textwidth,height=0.4\textheight}} & 
\mbox{\epsfig{figure=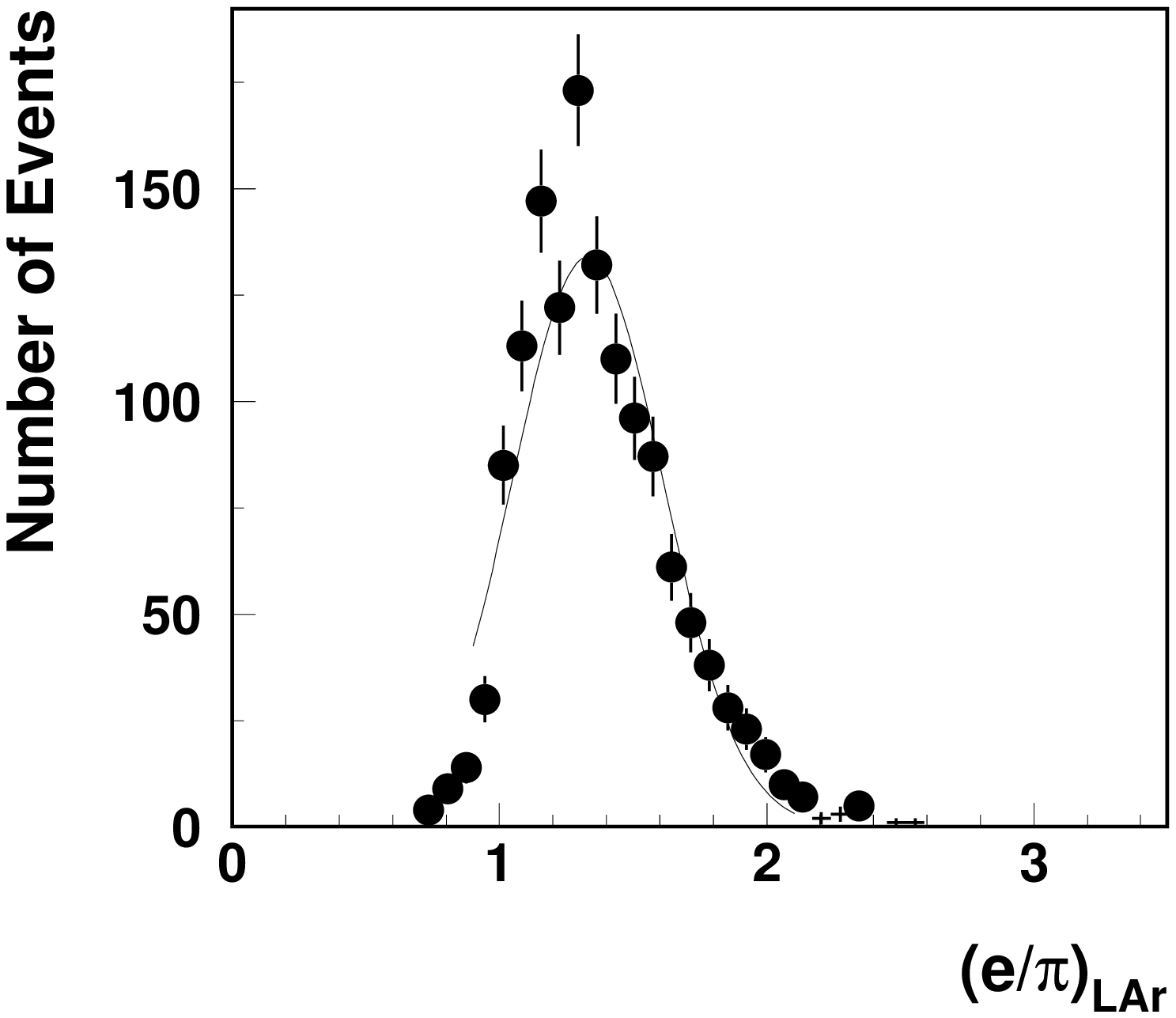,width=0.45\textwidth,height=0.4\textheight}}
\\
\end{tabular}
\end{center}
       \caption{The distributions of the $(e / \pi)_{LAr}$ ratio 
                for $E_{beam}$ = 10, 40  GeV
                (left column, up to down) 
                and $E_{beam}$ = 20, 50 GeV
                (right column, up to down). 
       \label{fv2}}
\end{figure*}

\begin{figure*}[tbph]
\begin{center}   
\begin{tabular}{cc}
\mbox{\epsfig{figure=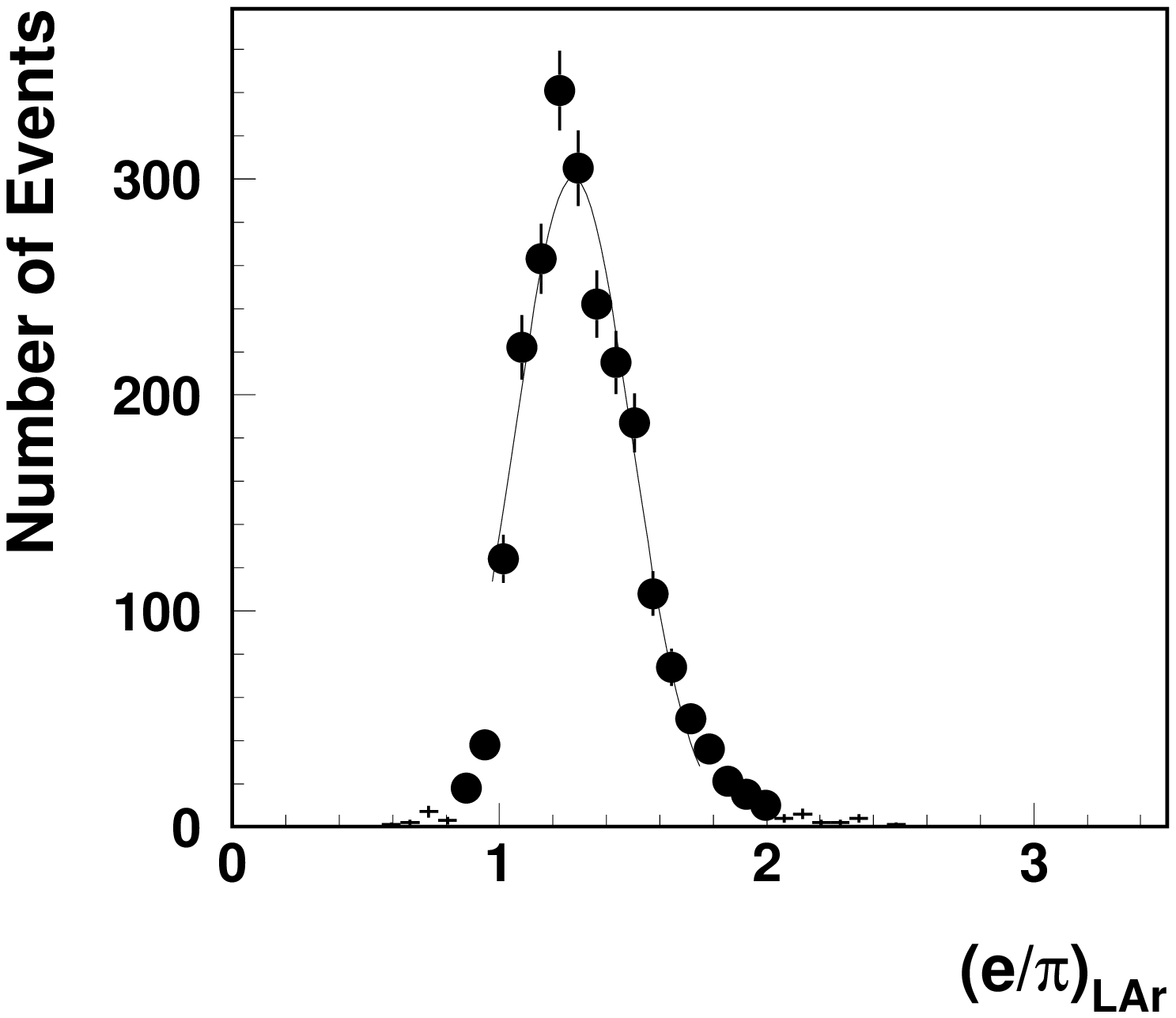,width=0.45\textwidth,height=0.4\textheight}} & 
\mbox{\epsfig{figure=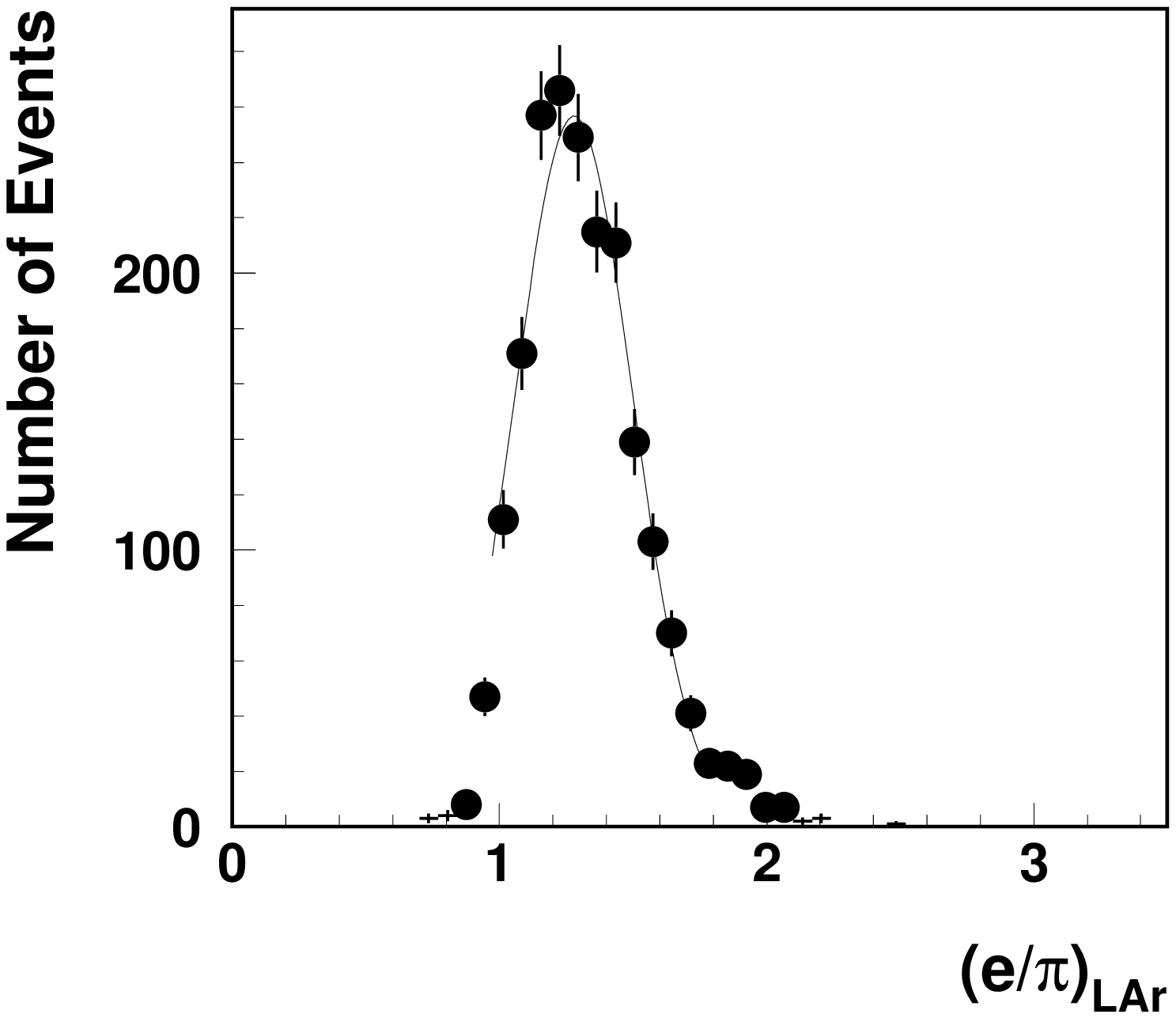,width=0.45\textwidth,height=0.4\textheight}} 
\\
\mbox{\epsfig{figure=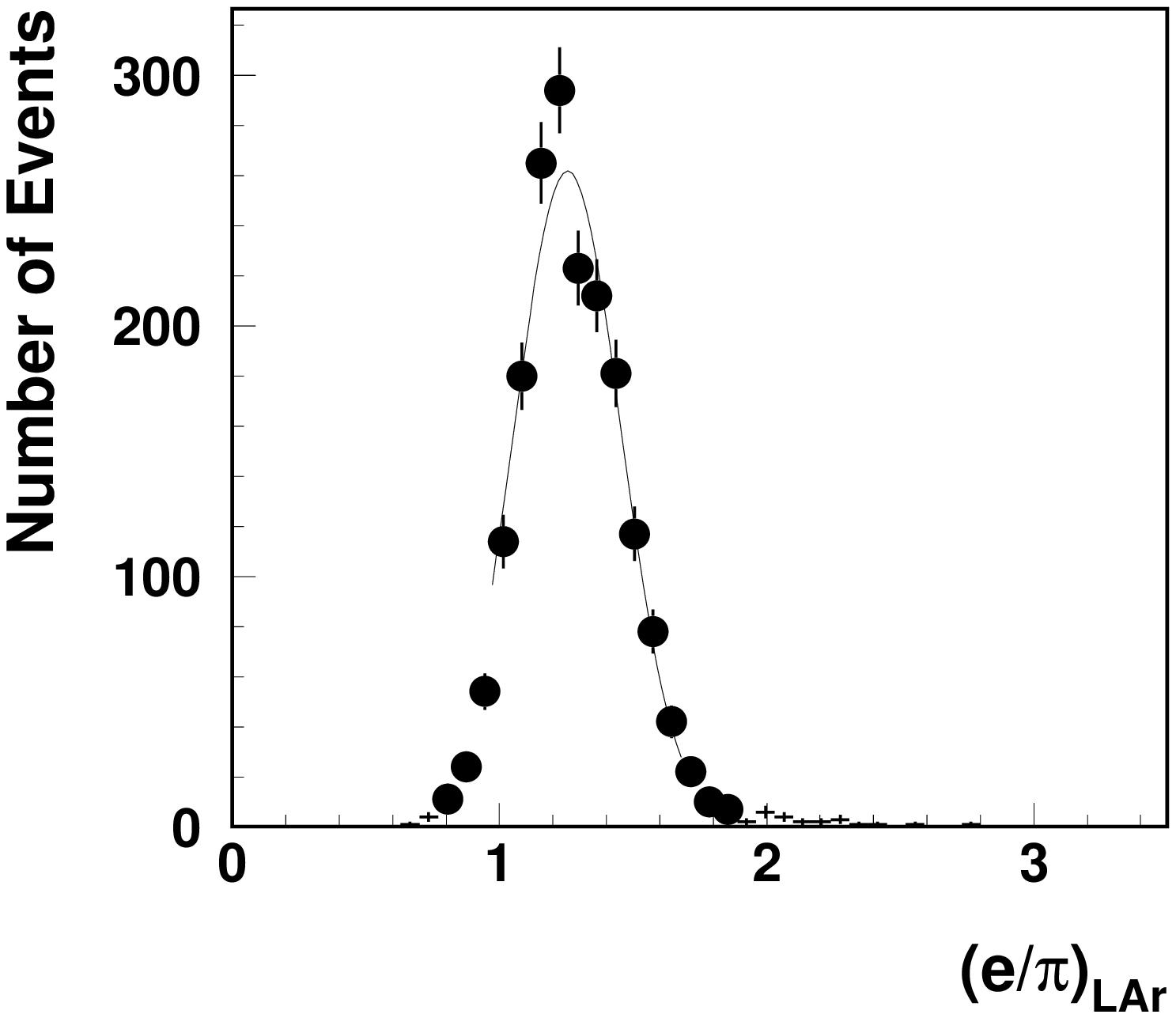,width=0.45\textwidth,height=0.4\textheight}} &
\mbox{\epsfig{figure=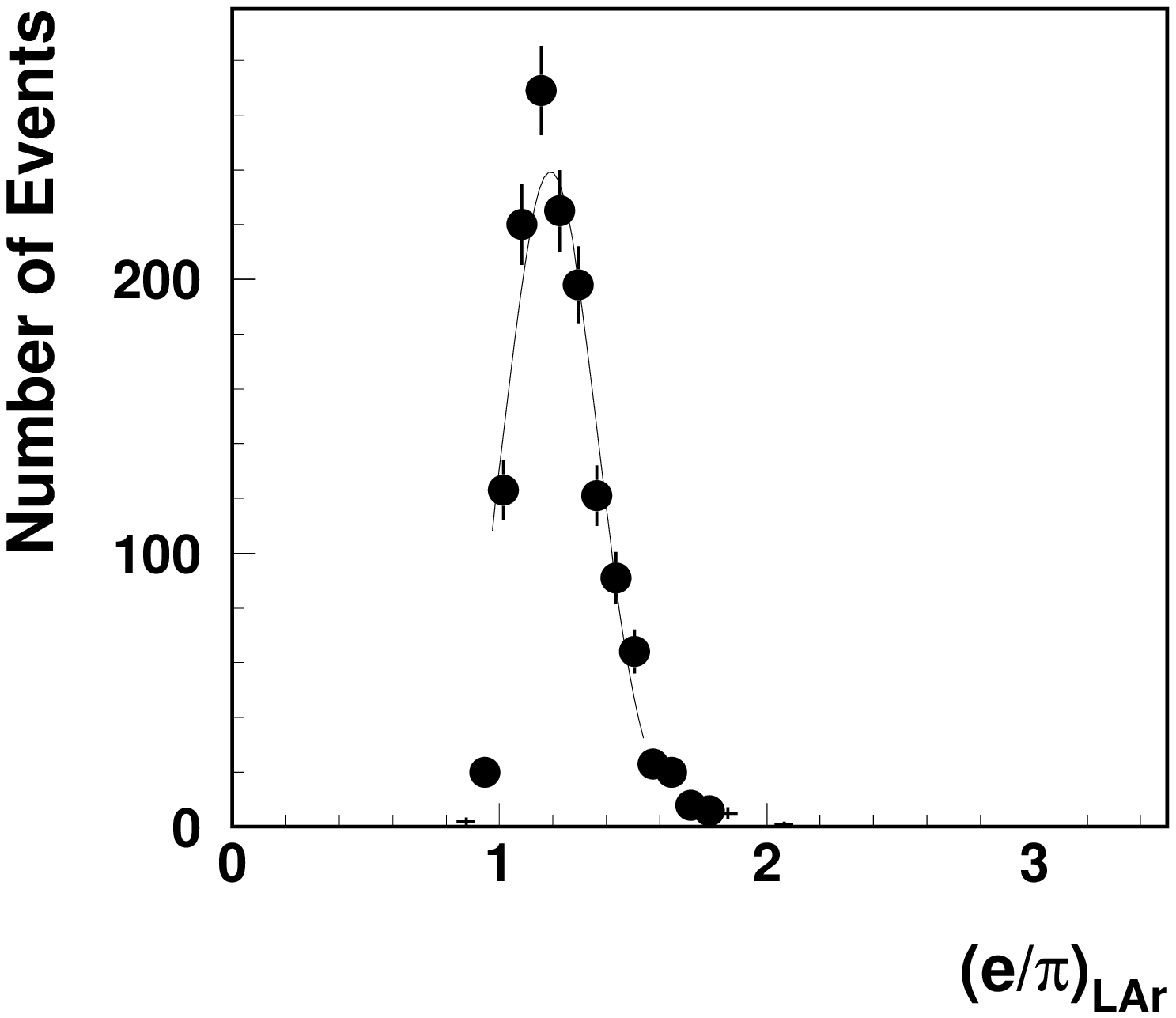,width=0.45\textwidth,height=0.4\textheight}} 
\\
\end{tabular}
\end{center}
       \caption{The distributions of the $(e / \pi)_{LAr}$ ratio 
                for $E_{beam}$ = 80, 150  GeV
                (left column, up to down) 
                and $E_{beam}$ = 100, 300  GeV
                (right column, up to down). 
       \label{fv2-1}}
\end{figure*}

\begin{figure*}[tbph]
\begin{center}   
\mbox{\epsfig{figure=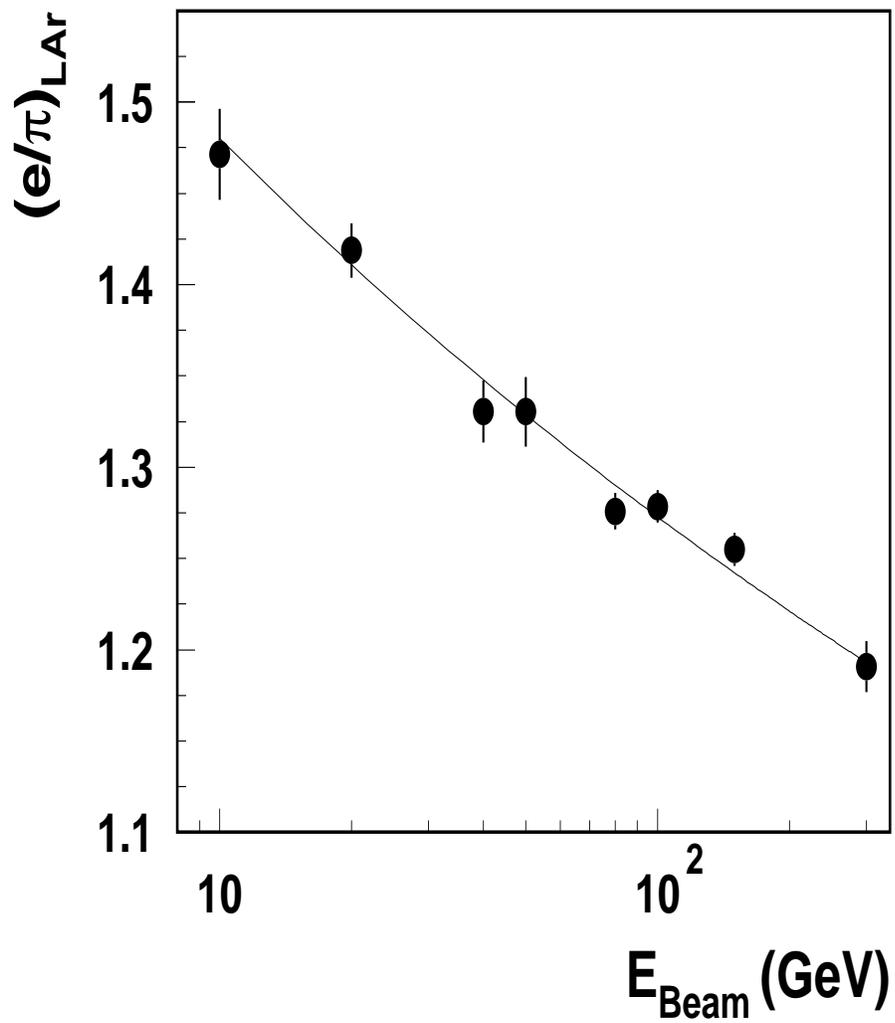,width=0.95\textwidth,height=0.85\textheight}}
\end{center}
       \caption{
	The mean values of the $(e/\pi)_{LAr}$ ratios as a function of the beam 
	energy.
        The line is the result of a fit of eq.\ (\ref{ev5}).
      \label{fv3}}
\end{figure*}

%%%%%%%%%%%%%%%%%%%%%%%%%%%%%%%%%%%%%%%%%%%%%%%%%%%%%%% 
\hspace{6mm}      
Figs.\ \ref{fv2} and \ref{fv2-1} 
show the distributions of the $(e/\pi)_{LAr}$ ratio derived
by formula (\ref{ev1}) for different energies.
The mean values of these distributions are given in Table~ \ref{tv1} and
shown in Fig.~\ref{fv3} as a function of the beam energy.
The fit of this distribution by the expression (\ref{ev5}) yields
$(e/h)_{LAr} = 1.74\pm0.04$ and $k_{LAr} = 0.108\pm0.004$
(${\chi^2} / NDF = 0.93$).
For the fixed value of the parameter $k_{LAr} = 0.11$ 
\cite{wigmans91} 
the result is $(e/h)_{LAr} = 1.77\pm0.02$ (${\chi^2} / NDF = 0.86$).
The quoted errors are the statistical ones obtained from the fit.
The systematic error on the $(e/h)_{LAr}$ ratio, which is a consequence of
the uncertainties in the input constants used in the equation (\ref{ev1}),
is estimated to be $\pm0.04$.

Wigmans  showed \cite{wigmans91} that the the $e/h$ ratio for non-uranium
calorimeters with high-Z absorber material is satisfactorily described
by the formula:
\begin{equation}
	\frac{e}{h}  = \frac{e/mip}{0.41 + 0.12\ n/mip}
\label{wig}
\end{equation}
in which $e/mip$ and $n/mip$ represent the calorimeter response to e.m.\
showers and to MeV-type neutrons, respectively.
These responses are normalized to the one for minimum ionizing particles.
The Monte Carlo calculated $e/mip$ and $n/mip$ values for the RD3 Pb-LAr
electromagnetic calorimeter are $e/mip = 0.78$ and $n/mip < 0.5$ leading to 
$(e/h)_{LAr} > 1.66$.
Our measured value of the $(e/h)_{LAr}$ ratio agrees with this prediction.  

There is the estimation of the $(e/h)_{LAr}$ ratio of $3.7 \pm 1.7$
for this electromagnetic compartment obtained in 
%20
\cite{stipcevic93}
on the basis of data from the combined lead-iron-LAr calorimeter
%21
\cite{gingrich95}.
This value  agrees with our value within errors.
But we consider their method as the incorrect one since 
for the determination of the $(e/\pi)_{LAr}$ ratios 
the calibration constants are used which have been obtained
by minimizing the energy resolution that leads to distortion
of the true $(e/\pi)_{LAr}$ ratios.

%%%%%%%%%%%%%%%%%%%%%
\section{Conclusions}
%%%%%%%%%%%%%%%%%%%%%
\hspace{6mm}
The method of extraction of the $e/h$ ratio,  the degree of 
non-com\-pen\-sa\-tion,  for the electromagnetic compartment
of the ATLAS barrel combined prototype calorimeter is suggested.
On the basis of the 1996 combined test beam data we have determined
this value which turned out to be equal to $1.74\pm0.04$ and 
agrees with the Monte Carlo prediction of Wigmans  that $e/h > 1.7$ for this 
LAr calorimeter.

%%%%%%%%%%%%%%%%%%%%%%%%%
\section{Acknowledgments}
%%%%%%%%%%%%%%%%%%%%%%%%%
\hspace{6mm}
This work is the result of the efforts of many people from the ATLAS
Collaboration.
The authors are greatly indebted to all Collaboration
for their test beam setup and data taking.
Authors are grateful Peter Jenni and Marzio Nessi for fruitful discussion
and support of this work. 
We are thankful Julian Budagov and Jemal Khubua for their attention and 
support of this work.
We are also thankful Illias Efthymiopoulos for giving the results of the 
Monte Carlo simulation, Irene Vichou and Marina Cobal for constructive 
advices and fruitful discussion.

%%%%%%%%%%%%%%%%%%%%%%%%%%
%  bibliography
%%%%%%%%%%%%%%%%%%%%%%%%%%

%%%%%%%%%%%%%%%%%%%%%

%\vspace*{\fill}

%%%%%%%%
% TABLES 
%%%%%%%%

%1
\begin{table}[tbph]
\begin{center}
\caption{	
		The  mean $(e / \pi)_{LAr}$ ratio    
        	as a function of the beam energy.}
\label{tv1}
\begin{tabular}{|c|c|}
\multicolumn{2}{l}{\mbox{~~~}} 		\\
\hline
$E_{beam} (GeV)$& $(e / \pi)_{LAr}$ 	\\ 
\hline
10  		& $1.471 \pm 0.025$     \\ 
\hline 
20  		& $1.419 \pm 0.015$     \\ 
\hline 
40  		& $1.331 \pm 0.017$     \\ 
\hline 
50  		& $1.330 \pm 0.019$    	\\ 
\hline 
80  		& $1.276 \pm 0.010$    	\\ 
\hline 
100 		& $1.278 \pm 0.009$ 	\\ 
\hline 
150 		& $1.255 \pm 0.009$ 	\\ 
\hline 
300 		& $1.191 \pm 0.014$ 	\\ 
\hline 
\end{tabular}
\end{center}
\end{table}

%\vspace*{\fill}

%%%%%%%%%
%Figures
%%%%%%%%%

%%%%%%%%%%%%%%%%%%%%%%%%%%%%%%%%%%%%%%%%%%%%%%%%%%%%%%%%%%%%%%%%%%%
%\newpage
\vspace*{\fill}
%1
%\clearpage

%\vspace*{\fill}
\newpage

%%%%%%%%%%%%%%%%%%%%%%%%%%%%%%%%%%%%%%%%%%%%%%%%%%%%%%%%%%%%%%%%%%%%%
%2
\clearpage

%%%%%%%%%%%%%%%%%%%%%%%%%%%%%%%%%%%%%%%%%%%%%%%%%%%%%%%%%%%%%%%%%%%%%%%%%
%3
\clearpage

%%%%%%%%%%%%%%%%%%%%%%%%%%%%%%%%%%%%%%%%%%%%%%%%%%%%%%%%%%%%%%%%%%%%%%%%%%%%
%4
\clearpage

%%%%%%%%%%%%%%%%%%%%%%%%%%%%%%%
%5
\clearpage

%%%%%%%%%%%%%%%%%%%%%%%%%%%%%%%
%6
\clearpage

%%%%%%%%%%%%%%%%%%%%%%%%%%%%%%%%%%%%%%%%%%%%%%%%%%%%%%%%%%%%%%%%%%%%%%%%%%%%
%7
\clearpage

%%%%%%%%%%%%%%%%%%%%%%%%%%%%%%%%%%%%%%%%%%%%%%%%%%%%
\end{document}